\documentclass{article}

\usepackage{arxiv}

\usepackage[utf8]{inputenc} 
\usepackage[T1]{fontenc}    
\usepackage{hyperref}       
\usepackage{url}            
\usepackage{booktabs}       
\usepackage{amsfonts}       
\usepackage{nicefrac}       
\usepackage{microtype}      
\usepackage{lipsum}
\usepackage{amsmath}
\usepackage{amssymb}
\usepackage{graphicx}
\graphicspath{ {./images/} }

\title{COVID-19 Detection Based on Blood Test Parameters using Various Artificial Intelligence Methods}

\author{
 Kavian Khanjani \\
  Department of Electrical Engineering\\
  K. N. Toosi University of Technology\\
  Tehran, Iran  \\
  \texttt{kaviankhanjani@email.kntu.ac.ir} \\
   \And
 Seyed Rasoul Hosseini \\
  Department of Electrical Engineering\\
  K. N. Toosi University of Technology\\
  Tehran, Iran \\
  \texttt{s.r.hosseini@email.kntu.ac.ir} \\
   \And
 Hamid Taheri \\
  Department of Electrical Engineering\\
  K. N. Toosi University of Technology\\
  Tehran, Iran \\
  \texttt{taheri.hamiid@gmail.com} \\
   \And
 Shahrzad Shashaani \\
  Department of Electrical Engineering\\
  K. N. Toosi University of Technology\\
  Tehran, Iran  \\
  \texttt{sh.shashaani@email.kntu.ac.ir} \\
   \And
 Mohammad Teshnehlab \\
  Department of Electrical Engineering\\
  K. N. Toosi University of Technology\\
  Tehran, Iran  \\
  \texttt{teshnehlab@eetd.kntu.ac.ir} \\
}

\begin{document}
\maketitle
\begin{abstract}
In 2019, the world faced a new challenge: a COVID-19 disease caused by the novel coronavirus, SARS-CoV-2. The virus rapidly spread across the globe, leading to a high rate of mortality, which prompted health organizations to take measures to control its transmission. Early disease detection is crucial in the treatment process, and computer-based automatic detection systems have been developed to aid in this effort. These systems often rely on artificial intelligence (AI) approaches such as machine learning, neural networks, fuzzy systems, and deep learning to classify diseases. This study aimed to differentiate COVID-19 patients from others using self-categorizing classifiers and employing various AI methods. This study used two datasets: the blood test samples and radiography images. The best results for the blood test samples obtained from San Raphael Hospital, which include two classes of individuals, those with COVID-19 and those with non-COVID diseases, were achieved through the use of the Ensemble method (a combination of a neural network and two machines learning methods). The results showed that this approach for COVID-19 diagnosis is cost-effective and provides results in a shorter amount of time than other methods. The proposed model achieved an accuracy of 94.09\% on the dataset used. Secondly, the radiographic images were divided into four classes: normal, viral pneumonia, ground glass opacity, and COVID-19 infection. These were used for segmentation and classification. The lung lobes were extracted from the images and then categorized into specific classes. We achieved an accuracy of 91.1\% on the image dataset. Generally, this study highlights the potential of AI in detecting and managing COVID-19 and underscores the importance of continued research and development in this field.
\end{abstract}


\section{Introduction}
Coronavirus disease 2019 (COVID-19) caused by the severe acute respiratory syndrome coronavirus 2 (SARS-CoV-2) \cite{r1, r2}, emerged in December 2019 and rapidly developed into a global outbreak \cite{r3, r4, r5, r6}. COVID-19 appears as an acute respiratory tract infection syndrome \cite{r8}. Patients infected with COVID-19 have a high mortality rate \cite{r9}. Because of the COVID-19 symptomatology, which showed a large number of asymptomatic \cite{r10}, the disease has become difficult to diagnose. The typical clinical signs of COVID-19 cases include fever, respiratory symptoms, pneumonia, decreased white blood cell (WBC) count, or decreased lymphocyte count \cite{r11, r12}, which are symptoms of the infection. In a more critical situation, this infection can lead to pneumonia, septic shock, SARS, organ failure, and death.

Since 2019, the virus has shown various mutations that have increased its pathogenicity. These changes have led to the creation of new species. One was the alpha-type mutation, which had a higher transmission rate than the original type. The beta version was identified in South Africa, which led to an increase in disease severity and mortality. A third species, called gamma, became common in Brazil, which had a higher transmission rate than the previous two species and reduced the effectiveness of vaccines. The most advanced species, lambda, was identified in India as having a higher transmission rate and increasing mortality and is the predominant species worldwide \cite{r14}. Respiratory problems spread faster in cases with COVID-19-induced pneumonia than in healthy instances \cite{r15}. Thus, tests to identify the SARS-CoV-2 virus play a key role in determining positive cases of this infection and thus curb the pandemic \cite{r16, r17}.

Several COVID-19 screening methods exist, such as chest computed tomography (CT), X-ray images, RT-PCR, and blood tests. Medical imaging, CT, is often used as a complementary examination in the diagnosis and management of COVID-19 \cite{r2, r18, r19}. This imaging interpretation played a key role in diagnosing and treating COVID-19 and monitoring disease progression \cite{r20, r21}. On the other hand, evidence of the disease in medical images may appear at least some days after the onset of symptoms, which leads to late detection and ultimately reduces the chances of patients’ recovery and preventing transmission \cite{r22}. Blood tests and radiocardiography (X-ray) are suitable diagnostic methods for COVID-19, and we have chosen to utilize data from both in our article. While blood tests have advantages over other methods such as X-rays, CT scans \cite{r23}, and PCR tests \cite{r24}, radiocardiography can complement them to improve accuracy and speed up diagnosis. Blood tests are cost-effective and provide faster results, even before symptoms appear, while radiocardiography can detect lung changes that may not be visible on blood tests. Additionally, blood tests do not have adverse side effects such as X-ray exposure.

However, other diagnostic methods have certain drawbacks. CT scans and X-rays, for example, can expose patients to harmful ionizing radiation, which can increase the risk of cancer. On the other hand, PCR tests can take longer to yield results and may not detect the virus in the early stages of infection. Furthermore, these tests rely on the presence of symptoms, which means that asymptomatic carriers may not be identified until it is too late. Artificial intelligence can be integrated further to enhance the accuracy and efficiency of these diagnostic methods, as numerous studies have already explored.

\section{Related Works}
Machine learning techniques have been employed to detect COVID-19 from lung CT scans with 90\% sensitivity \cite{r26, r27}. A deep learning technique was also developed to extract the graphical characteristics of COVID-19 from CT images to make diagnosis quicker and more precise than pathogenic testing to save critical time \cite{r17, r28}. Several deep learning architectures are deployed to detect COVID-19, such as ResNet, Inception, Google Net, etc. All these approaches can detect subjects who suffer from pneumonia, and it is hard to decide the source of pneumonia \cite{r17}. In addition, studies on the uncertainty of the methods implemented by deep learning to diagnose the disease from CT images are done \cite{r29, r30, r31}. Using X-ray images is a cheaper and easier way than CT. However, lower accuracy and quality of X-ray images can lead to misdiagnosis and a worse epidemic \cite{r31, r32}. In other studies, by using deep learning with different architectures such as ResNet50 \cite{r33, r34}, VGG \cite{r35}, Mobile Net \cite{r36}, etc., they were able to find acceptable results for correct diagnosis \cite{r17}. Due to the different symptoms, the polymer chain reaction (PCR) is considered a standard method for screening suspicious cases \cite{r17, r37, r39} Using blood test parameters, each parameter's effect on the disease's development is determined to be much faster and less expensive than other methods \cite{r40}. In addition, using different classification methods, models with an accuracy between 82\% and 86\% have been obtained \cite{r39}.

These studies highlight the potential of combining blood test parameters and X-ray imaging with various artificial intelligence methods for COVID-19 detection. While further research is needed to validate the efficacy of these methods in clinical practice, the results suggest that they could be promising tools for improving diagnostic accuracy and reducing false positives. The article is organized into several sections. Section 3 explains the data and preprocessing procedures. In section 4, the methods used for classification are introduced. Section 5 presents implementation details and results, while section 6 discusses related topics.

\section{Data Description}
In this work, the dataset includes Blood test data of patients adopted from San Raphael Hospital, Italy.

\subsection{Blood dataset (San Raphael Hospital)}
The provided information in this section contains Alanine Amino Transferase (ALT), Aspartate Aminotransferase (AST), Lactate Dehydrogenase (LDH), Urea, White blood cells (WBC), Red Blood Cells (RBC), Platelet, Monocyte, Lymphocyte, and Neutrophil. The dataset has two classes, COVID-19 and non-COVID, identified by labels 1 and 0, respectively. This dataset was adopted from San Raphael Hospital and is available on link \footnote{\href{https://zenodo.org/record/4081318\#.YbyAQ2hBxPZ}{https://zenodo.org/record/4081318}}. The number of cases includes 505 blood test samples with a label 1 for patients with COVID-19 and 410 samples with a label of 0 for people without this disease \cite{r42}. More details are shown in Table \ref{table1}

\begin{table}[ht]
\centering
\caption{Blood test data (San Raphael Hospital) details}
\label{table1}
\begin{tabular}{c|cc}
Disease      & \multicolumn{1}{c|}{Number of samples} & Target label \\ \hline
Non-COVID-19 & 410                                    & 0            \\
COVID-19     & 505                                    & 1        
\end{tabular}
\end{table}

\subsection{Classification Data}
The dataset contains four classes of radiography images. Categories include patients with COVID-19, normal, Viral pneumonia, and grand glass opacity. The details are given in Table \ref{table2}.

\begin{table}[ht]
\centering
\caption{Classification data details}
\label{table2}
\begin{tabular}{c|cc}
Type of Image   & \multicolumn{1}{c|}{Number of Images} & Label \\ \hline
Normal          & 3000                                  & 0     \\
Viral Pneumonia & 1345                                  & 1     \\
Opacity         & 3000                                  & 2     \\
COVID-19        & 3000                                  & 3     
\end{tabular}
\end{table}

The images are presented in grayscale and with the original version of $299 \times 299$ in PNG format. Images are resized to $256 \times 256$ to apply to the classifier. Before performing other processing, the data sets are Normalized between zero and one. 

\subsection{Segmentation Data}
This section's data for segmenting lung lobes includes $3500$ images with $3500$ corresponding masks. Because the images' dimensions differ, the masks have been resized to $256 \times 256$ to match the images.

\section{Methodology}
This section summarizes the significant methods and their mathematical basis in separate sections.

\subsection{Proposed Method}
\begin{figure}
  \centering
  \includegraphics[width=11cm, height=6cm]{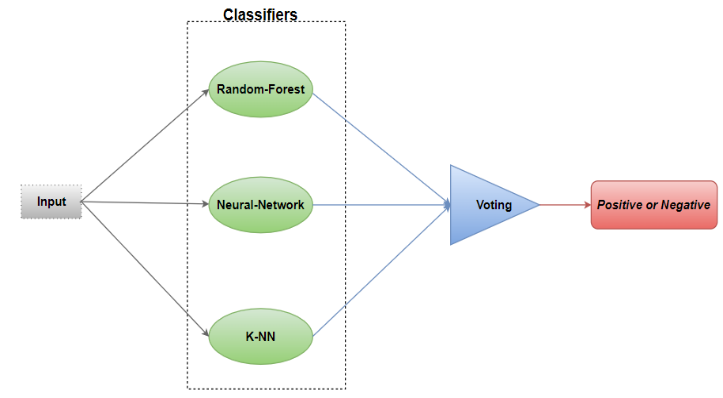} 
  \caption{Overall structure of Ensemble method.}
  \label{fig0}
\end{figure}

If a compound of classifiers is used for classification, an ensemble method will be created. One combination method, for example, is voting among the classifiers used, and whichever group gets the most votes will be used as the overall output. In this work, a mixture of categorizers, including multi-layer perceptron, k-nearest-neighborhood, and random forest, has been used. The overall structure of the ensemble method is shown in Figure \ref{fig0}.

\subsection{Adaptive Neural Fuzzy Inference System}
Adaptive Neural Fuzzy Inference System (ANFIS) is a hybrid model consisting of fuzzy logic that uses the fuzzy Takagi-Sugeno model and an artificial neural network designed to determine the behavior of highly complex dynamic systems. An adaptive fuzzy network can interpret and represent prior knowledge. It creates several if-then fuzzy rules with appropriate membership functions to generate the input's correct output. The capability of back-propagation in neural networks can be used to train fuzzy rules' membership functions. The ANFIS structure consists of five layers, namely, fuzzification layer, product layer, normalized layer, de-fuzzification layer, and output layer. The mathematical explanation of ANFIS layers is summarized below:

Layer 1: Each node $ij$ generates the membership function (MF) $\mu_{i,j}$ for the corresponding crisp input $x_i$ based on Eq.\ref{eq1}.

\begin{equation}
O_{i,j}^{1} = \mu_{A_{i,j}(x_i)}, \quad i = 1, ..., n, \quad j = 1, ..., p
\label{eq1}
\end{equation}

In this study, the MF type is Gaussian functions.

\begin{equation}
\mu_{A_{i,j}}(x) = \exp \left( - \left(\frac{(x - C_{i,j})}{a_{i,j}}\right)^{2b_{i,j}} \right).
\label{eq2}
\end{equation}

$a_{i,j}$, $b_{i,j}$, and $c_{i,j}$ are trainable parameters known as premise parameters.

Layer 2: Each node $r$ in this layer calculates the firing strength of the rule $r$ following.

\begin{equation}
O_{r}^{2} = w_{r} = \mu_{A_{1,j_{l_1}}}(x)  \ldots  \mu_{A_{1,j_{l_n}}}(x), \quad r = 1, ..., R, \quad i = 1, ..., p
\label{eq3}
\end{equation}

Layer 3: The nodes of this layer normalize the firing strength of rule $r$ from the previous layer in Eq. \ref{eq4}.

\begin{equation}
O_{r}^{3} = \bar{w}_{r} = \frac{w_{r}}{\sum_{k=0}^{r} w_{k}}
\label{eq4}
\end{equation}

Layer 4: Output of this layer indicates the contribution of each node $r$ in output based on Eq. \ref{eq5}.

\begin{equation}
O_{r}^{4} = \bar{w}_{r}f_r = \bar{w}_{r} \left( \sum_{i=1}^{n} c_{r,i}x_i + e_r \right)
\label{eq5}
\end{equation}

Parameters $c_{r, i}$, and $e_r$ are trainable parameters known as consequent parameters.

\textbf{Layer 5:} The output of this layer represents the total output as the summation of all rules’ results of incoming inputs given in Eq. \ref{eq6}.

\begin{equation}
O_r^5 = \sum_{r=1}^{R} \bar{W_r} f_r = \frac{\sum_{r=1}^{R} W_r f_r}{\sum_{r=1}^{R} W_r}
\label{eq6}
\end{equation}

\begin{figure}
  \centering
  \includegraphics[width=15cm, height=8cm]{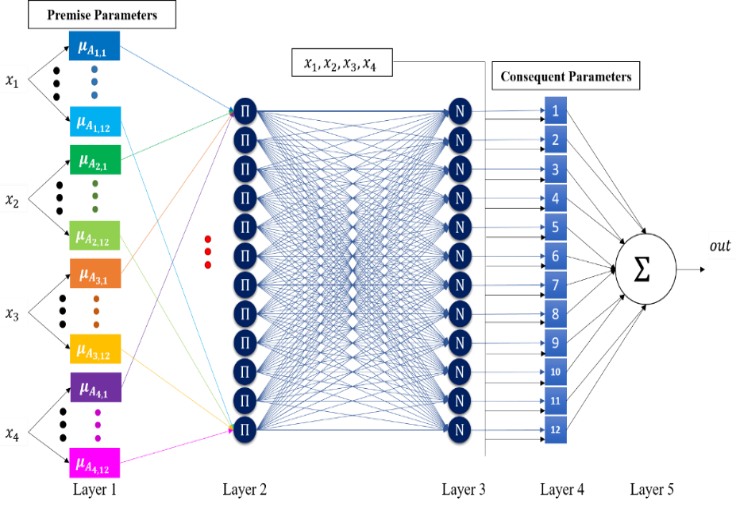} 
  \caption{Structure of Adaptive Neural Fuzzy Inference System.}
  \label{fig1}
\end{figure}

The overall structure of the ANFIS model is shown in Figure \ref{fig1} \cite{r41}.

\subsection{Data Preprocessing}
\begin{figure}
  \centering
  \includegraphics[width=15cm, height=9cm]{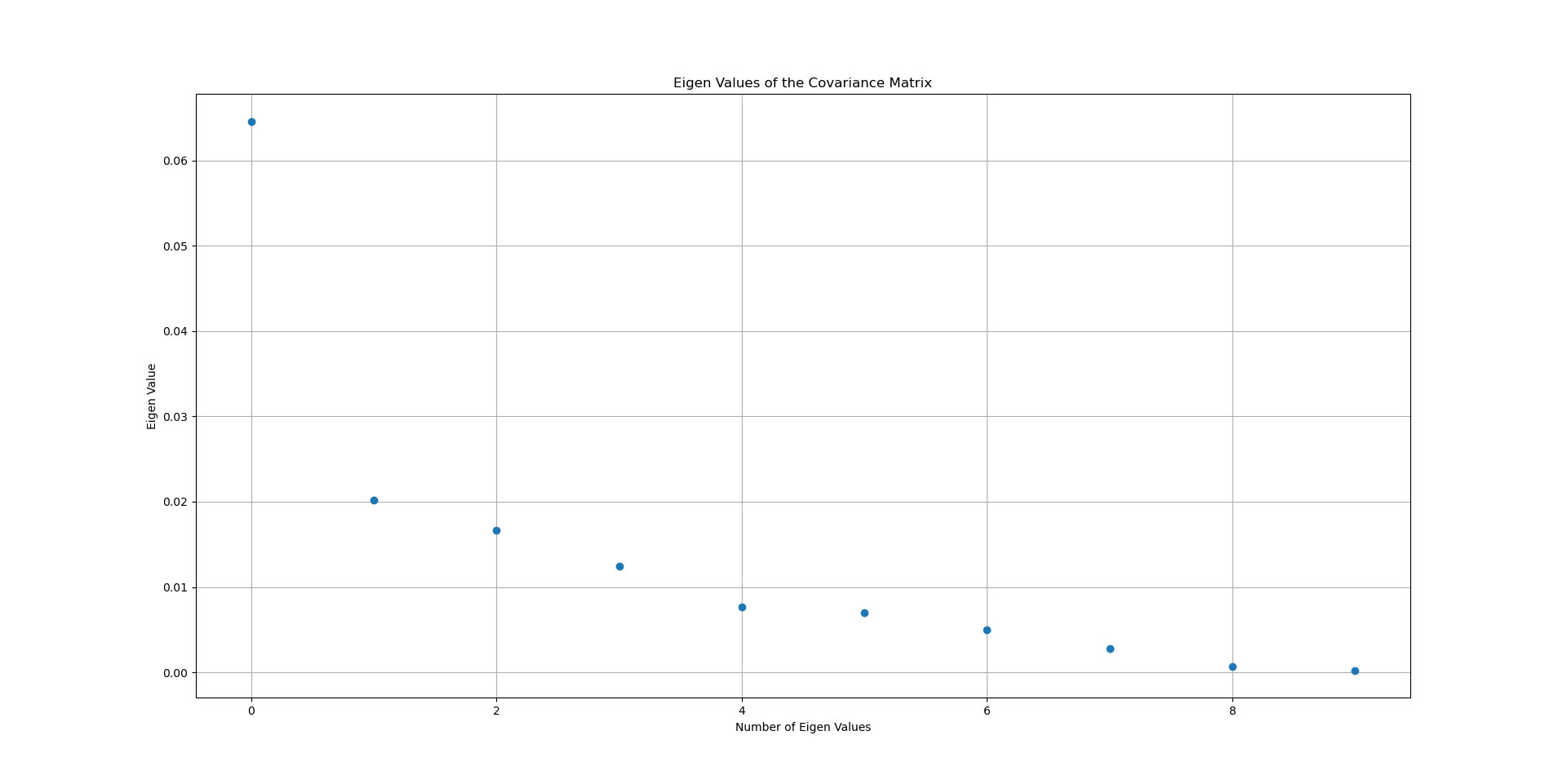} 
  \caption{Covariance matrix eigenvalues-Blood test data.}
  \label{fig2}
\end{figure}

Two preprocessing steps are performed to apply blood data as input to the designed classifiers. At first, information is normalized on the scale of [0, 1]. Then, a statistics function is calculated to get knowledge about the data distribution. The covariance matrix of datasets is a criterion for determining the data correlation and their statistical distribution. The covariance matrix's eigenvalues show the data's scatter in each direction. The graph of eigenvalues of the covariance matrix is shown in Figure \ref{fig2}. According to this Figure \ref{fig2}, the highest difference between eigenvalues is negligible. Therefore, the data is significantly distributed in all dimensions, and omitting each feature will reduce classification accuracy.

\begin{figure}
  \centering
  \includegraphics[width=17.5cm, height=17.5cm]{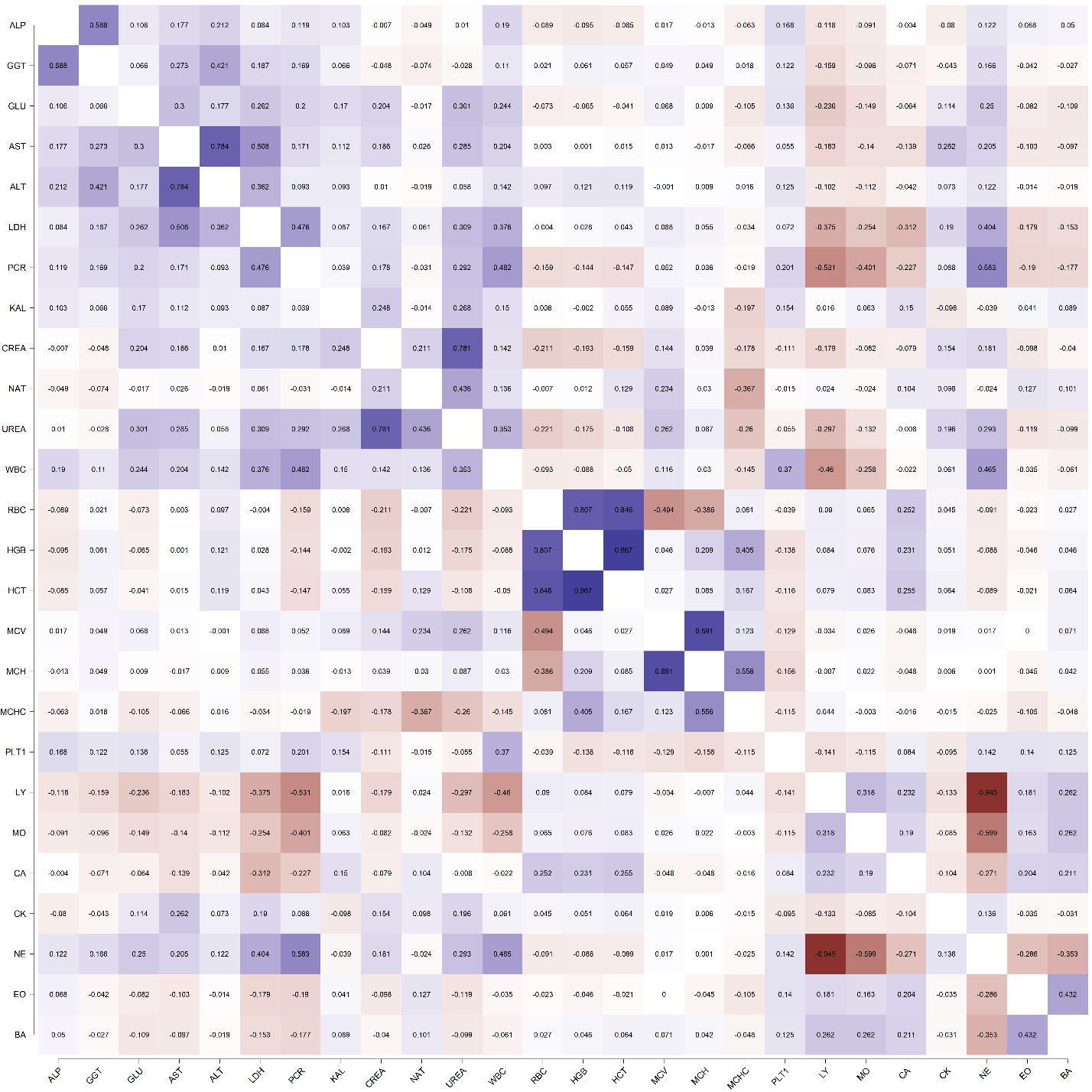} 
  \caption{Correlation Matrix for San Raphael Hospital data.}
  \label{fig3}
\end{figure}

Among 26 blood parameters adopted from San Raphael Hospital, ten features are extracted from the main data source, following the correlation coefficient Pearson. The correlation matrix is shown in Figure \ref{fig3}.

\subsection{Image preprocessing}
Various mappings can be applied to improve the contrast of the images. The changes mentioned in the histogram pot alter the images according to the user's purpose.

The histogram equalization transform improves the contrast of the image. It changes the brightness intensity of the pixels so that they are uniformly distributed in the defined range of values. In the Eq. \ref{eq7}, $h(x)$ represents the histogram equation.

\begin{equation}
h(x) = n_x \quad; x = 0, ..., L-1
\label{eq7}
\end{equation}

So $x$ specifies the brightness of each pixel, and $L$ is the maximum brightness. The $n_x$ parameter indicates the number of pixels with $x$ brightness.
The probability function of the brightness intensity values is considered as described in the Eq. \ref{eq8}.

\begin{equation}
p(x_k) = \frac{n_k}{MN} \quad; k = 0, \ldots, L-1
\label{eq8}
\end{equation}

The histogram balancing mapping is determined using the Eq. \ref{eq9}.

\begin{equation}
s_k = T(x_k) = (L-1) \sum_{j=0}^{k} p(x_j) = \frac{(L-1)}{MN} \sum_{j=0}^{k} n_j
\label{eq9}
\end{equation}

\begin{figure}
  \centering
  \includegraphics[width=14.5cm, height=10.5cm]{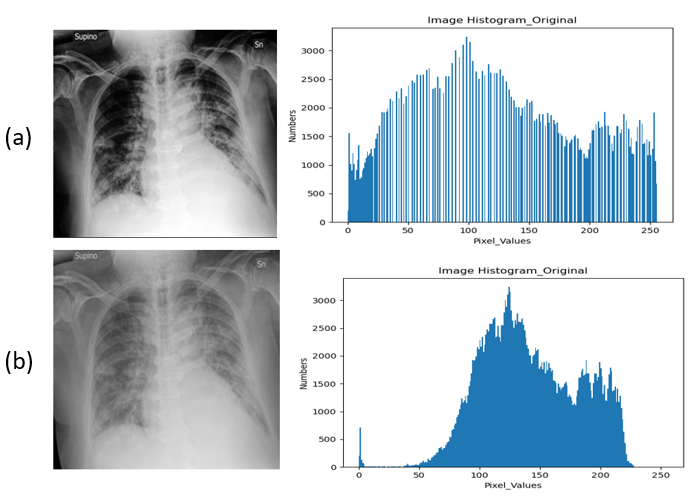} 
  \caption{Original, mapped images and corresponding histograms}
  \label{fig4}
\end{figure}

Figure \ref{fig4} shows the original, mapped images and corresponding histogram plots.

\begin{figure}
  \centering
  \includegraphics[width=16.5cm, height=10.5cm]{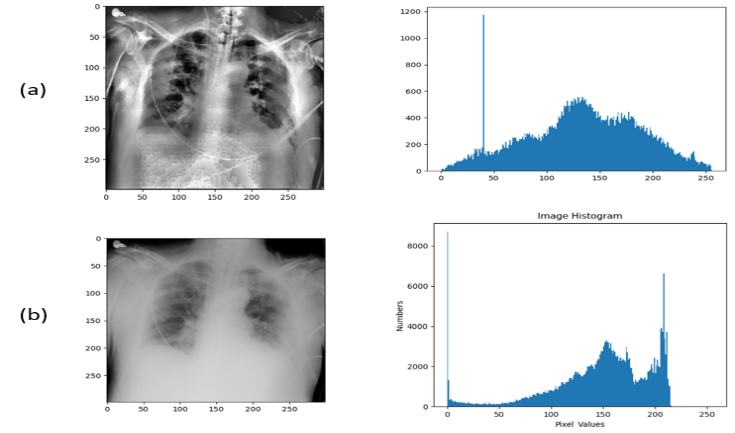} 
  \caption{Transformation of the X-ray image}
  \label{fig5}
\end{figure}

The contrast-limited adaptive histogram equalization transform works similarly to the histogram equalization method. This method prevents noises that may occur in the histogram balancing method. It divides the image matrix into smaller dimensions and performs equalization on each. Figure \ref{fig5} shows the transformation of the X-ray image.

The grad-CAM method is one of the generalization approaches of CAM. This method can be extended to various structures, such as categorizing images, captioning images, and answering image questions. Like the class activation map, the weights corresponding to each feature map are calculated. Then, the RELU activation function is applied to their linear combination to prevent the pixel values from becoming negative. The equations of the Grad-CAM method are according to \ref{eq10} and \ref{eq11}.

\begin{equation}
\alpha_k^c = \frac{1}{Z} \sum_i \sum_j \frac{\partial y^c}{\partial A_{ij}^k}
\label{eq10}
\end{equation}

\begin{equation}
L_{\text{Grad-CAM}}^c = \text{ReLU} \left( \sum_k \alpha_k^c A^k \right)
\label{eq11}
\end{equation}

The parameter $\alpha_k^c$ is the weight corresponding to the feature map $k$ and the score related to the class $c$. $A_{ij}^k$ represents the $ij$ domain of the feature map $A^k$. The $\frac{1}{Z}$ parameter corresponds to the global maximum integration layer (average feature matrices). Finally, the heat map is determined by the parameter $L_{\text{Grad-CAM}}^c$.

\subsection{Loss Functions}
The loss function used in autoencoder training is the mean square error, which is shown in \ref{eq12}. Another part of the loss function is the regularization term \ref{eq13}, which restricts the values of adaptive parameters, and the total function includes the summation of these two parts that are available in \ref{eq14}. 

\begin{equation}
J_{\text{mse}} = \frac{1}{M} \sum_{d=1}^{M} \left\lVert x^d - \hat{x}^d \right\rVert^2
\label{eq12}
\end{equation}

\begin{equation}
J_{\text{Regularization}} = \frac{\lambda}{2} \sum_{l=1}^{n_l-1} \sum_{i=1}^{S_l} \sum_{j=1}^{S_{l}+1} \left( w_{ij}^{l} \right)^2
\label{eq13}
\end{equation}

\begin{equation}
J_{\text{Total}} = J_{\text{mse}} + J_{\text{Regularization}} = \frac{1}{M} \sum_{d=1}^{M} \frac{1}{2} \left\lVert x^d - \hat{x}^d \right\rVert^2 + \frac{\lambda}{2} \sum_{l=1}^{n_l-1} \sum_{i=1}^{S_l} \sum_{j=1}^{S_{l}+1} \left( w_{ij}^{l} \right)^2
\label{eq14}
\end{equation}

$x^d$ is the input vector, $\hat{x}^d$ is an estimation vector (AE output) and batch size is $M$. $n_l$ shows the number of network layers, $S_l$ is the number of neurons in the $l$th layer and $w_{ij}^{l}$ demonstrates the weight between $i$th neuron in the $l+1$th layer and $j$th neuron in the $l$th layer.

\section{Implementation details}
In both experiments performed on the data, the 5-fold validation method is used. Thus, the normalized data is divided into five equal parts, and in each step, 80\% of the data are for training and 20\% for testing are considered. The results in each folder are presented separately, as well as the average. In this way, the performance of the designed categorizer regarding all data is evaluated.

\subsection{Experiments on Blood dataset (San Raphael Hospital)}
This data, with ten dimensions, has been classified using several methods. ANFIS model is trained with a learning rate of 0.001 and a number of epochs of 4000. The different number of rules and optimizers apply, and 14 and Adam optimizers obtain the best result. Besides, machine learning methods that classified blood test data in the previous section were used again for this data. Another approach is the stacked autoencoder, which reduces the dimensions of the input vector and extracts the main features. The number of inputs for the first AE is ten, and the number of neurons in the encoder layer is 20. Activation functions are leaky ReLU, and the coefficients of the learning rate, momentum, and regularizations are 0.008, 0.9, and 0.001, respectively. The training process is based on batch learning with 250 epochs, and the batch size is 128. The number of neurons in the encoder layer of the second autoencoder is 15, and for the third one is 8. Other details are similar to the first layer. The extracted features are applied to the last layer with two neurons using the SoftMax function.
After designing a stacked structure, the whole network is adopted based on the global training method. In this part, the optimizer is Adam, and the cross-entropy loss function is used. Its equation is shown in \ref{eq15}. 

\begin{equation}
J(y, \hat{y}) = -\sum_{i=1}^{k} y_i \log \hat{y}_i
\label{eq15}
\end{equation}

$\hat{y}_i$ is the classifier's estimation, and if it is the same as the target label, $y_i$ becomes 1; otherwise, this value is 0.

The results are represented by confusion matrices, which are an average of 5-folds, and the accuracy, precision, recall, and f1-score criteria given in \ref{eq16}, \ref{eq17}, \ref{eq18}, \ref{eq19} and \ref{eq20},, which determine the results in each fold.

\begin{equation}
\text{Accuracy} = \frac{\text{True Negative} + \text{True Positive}}{\text{True Negative} + \text{True Positive} + \text{False Positive} + \text{False Negative}}
\label{eq16}
\end{equation}

\begin{equation}
\text{Recall} = \frac{\text{True Positive}}{\text{True Positive} + \text{False Negative}}
\label{eq17}
\end{equation}

\begin{equation}
\text{Precision} = \frac{\text{True Positive}}{\text{True Positive} + \text{False Positive}}
\label{eq18}
\end{equation}

\begin{equation}
\text{F1-Score} = \frac{2 \times (\text{Recall} \times \text{Precision)}}{\text{Recall} + \text{precision}}
\label{eq19}
\end{equation}

\begin{equation}
\text{Inverse Recall} = \frac{\text{True Positive}}{\text{True Positive} + \text{False Negative}}
\label{eq20}
\end{equation}

\section{Results}
In this section, the outcomes of simulations are displayed. The results of each dataset are given in detail in two separate parts.
In this section, the outcomes of simulations are displayed. The results of each dataset are given in detail in two separate parts.

\subsection{Blood dataset (San Raphael Hospital)}
\begin{figure}
  \centering
  \includegraphics[width=11.5cm, height=8.5cm]{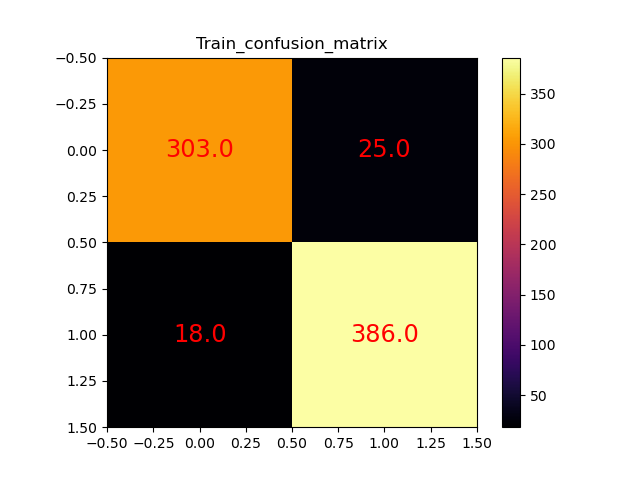} 
  \caption{Confusion matrix of train data on Blood test samples for San Raphael Hospital with Ensemble method.}
  \label{fig6}
\end{figure}

\begin{figure}
  \centering
  \includegraphics[width=11.5cm, height=8.5cm]{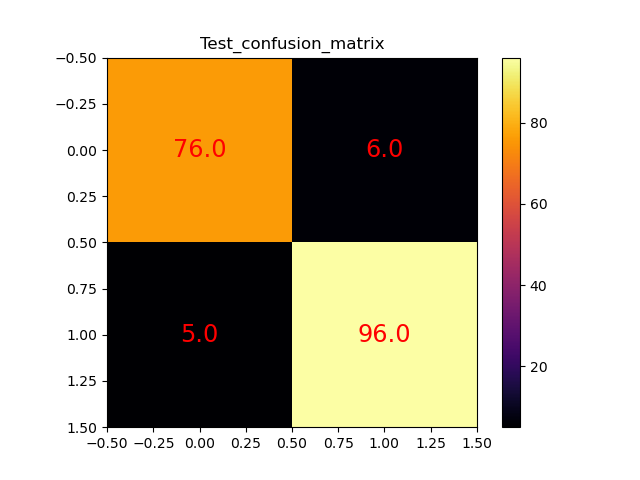} 
  \caption{Confusion matrix of test data on Blood test samples for San Raphael Hospital with Ensemble method.}
  \label{fig7}
\end{figure}

In this section, the results and parameters of the evaluation are similar to the previous section, except that the specifications and dimensions of the data are different. The best results are acquired from the Ensemble method. Element 11 in the confusion matrix is True Negative which shows the number of non-COVID persons that identified correctly. Element 2-2, True positive, shows the number of patients who have COVID-19 precisely. Figure \ref{fig6} and Figure \ref{fig7} are the average train and test confusion matrices of the Ensemble Classifier, respectively. The details of each fold of this procedure are specified in Table \ref{table3} and \ref{table4}.

\begin{table}[ht]
\centering
\caption{Evaluation parameters of each fold of train data-Blood test samples- San Raphael Hospital-Ensemble}
\label{table3}
\begin{tabular}{c|cccccc}
Train     & Fold-1 & Fold-2 & Fold-3 & Fold-4 & Fold-5 & Average \\ \hline
Accuracy  & 0.948  & 0.942  & 0.934  & 0.945  & 0.934  & 0.940   \\
Precision & 0.943  & 0.941  & 0.929  & 0.949  & 0.933  & 0.939   \\
Recall    & 0.962  & 0.955  & 0.952  & 0.954  & 0.947  & 0.954   \\
F1-Score  & 0.953  & 0.948  & 0.940  & 0.951  & 0.940  & 0.946  
\end{tabular}
\end{table}

\begin{table}[ht]
\centering
\caption{Evaluation parameters of each fold of test data-Blood test samples- San Raphael Hospital-Ensemble}
\label{table4}
\begin{tabular}{c|cccccc}
Test      & Fold-1 & Fold-2 & Fold-3 & Fold-4 & Fold-5 & Average \\ \hline
Accuracy  & 0.912  & 0.934  & 0.967  & 0.923  & 0.967  & 0.940   \\
Precision & 0.923  & 0.932  & 0.980  & 0.895  & 0.961  & 0.938   \\
Recall    & 0.923  & 0.950  & 0.963  & 0.955  & 0.980  & 0.954   \\
F1-Score  & 0.923  & 0.941  & 0.971  & 0.924  & 0.971  & 0.946 
\end{tabular}
\end{table}

The results of fuzzy systems with different numbers of rules are available in Table \ref{table5}. Table \ref{table6} and \ref{table7} are general comparisons of methods for blood test data of San Raphael hospital, which presents the percentages as the average of the folds.

\begin{table}[ht]
\centering
\caption{Fuzzy result with the number of rules 10, 12, 14, 16, and 18 respectively- San Raphael Hospital}
\label{table5}
\begin{tabular}{c|cccc|cccc}
      & \multicolumn{4}{c|}{Test}                & \multicolumn{4}{c}{Train}                \\ \hline
Classifiers & Accuracy & Precision & Recall & F1-Score & Accuracy & Precision & Recall & F1-Score \\ \hline
Model\_1 (10 Rules)    & 82\%     & 82\%      & 84\%   & 83\%     & 84\%     & 85\%      & 87\%   & 86\%     \\
Model\_2 (12 Rules)    & 80\%     & 78\%      & 84\%   & 81\%     & 85\%     & 86\%      & 87\%   & 87\%     \\
Model\_3 (14 Rules)    & 83\%     & 82\%      & 86\%   & 84\%     & 90\%     & 90\%      & 92\%   & 91\%     \\
Model\_4 (16 Rules)    & 85\%     & 86\%      & 87\%   & 87\%     & 89\%     & 93\%      & 87\%   & 90\%     \\
Model\_5 (18 Rules)    & 84\%     & 87\%      & 82\%   & 85\%     & 87\%     & 88\%      & 89\%   & 89\%  
\end{tabular}
\end{table}

\begin{table}[ht]
\centering
\caption{Comparison of classifiers based on evaluation parameters-Blood test data- San Raphael Hospital-Ensemble-TRAIN}
\label{table6}
\begin{tabular}{c|cccc}
                      & \multicolumn{4}{c}{Train}                 \\ \hline
Classifiers           & Accuracy & Precision & Recall  & F1-Score \\ \hline
K-NN                  & 99.11\%  & 99.90\%   & 99.94\% & 99.92\%  \\
Random-Forest         & 99.83\%  & 99.89\%   & 99.80\% & 99.85\%  \\
MLP                   & 81.69\%  & 83.62\%   & 83.46\% & 83.41\%  \\
Fuzzy Network         & 90\%     & 90\%      & 90\%    & 90\%     \\
SVM (kernel: Linear)  & 76.20\%  & 79.96\%   & 75.93\% & 77.88\%  \\
SVM (kernel: RBF)     & 80.05\%  & 83.35\%   & 79.79\% & 81.53\%  \\
SVM (kernel: Sigmoid) & 51.44\%  & 54.74\%   & 69.02\% & 61.05\%  \\
SAE                   & 79.04\%  & 81.68\%   & 79.94\% & 80.80\%  \\
\textbf{Ensemble}              & \textbf{94.09}\%  & \textbf{93.95}\%   & \textbf{95.44}\% & \textbf{94.69}\% 
\end{tabular}
\end{table}

\begin{table}[ht]
\centering
\caption{Comparison of classifiers based on evaluation parameters-Blood test data- San Raphael Hospital-Ensemble-TEST}
\label{table7}
\begin{tabular}{c|cccc}
                      & \multicolumn{4}{c}{Test}                 \\ \hline
Classifiers           & Accuracy & Precision & Recall  & F1-Score \\ \hline
K-NN                  & 67.65\%  & 70.19\%   & 72.06\% & 71.01\%  \\
Random-Forest         & 78.46\%  & 80.46\%   & 80.71\% & 80.50\%  \\
MLP                   & 76.72\%  & 78.78\%   & 79.80\% & 78.98\%  \\
Fuzzy Network         & 83\%     & 83\%      & 83\%    & 83\%     \\
SVM (kernel: Linear)  & 74.86\%  & 78.42\%   & 75.32\% & 76.78\%  \\
SVM (kernel: RBF)     & 76.93\%  & 80.74\%   & 76.51\% & 78.55\%  \\
SVM (kernel: Sigmoid) & 52.24\%  & 55.44\%   & 69.87\% & 61.64\%  \\
SAE                   & 78.25\%  & 80.76\%   & 79.68\% & 80.16\%  \\
\textbf{Ensemble}              & \textbf{94.09}\%  & \textbf{93.88}\%   & \textbf{95.44}\% & \textbf{94.64}\% 
\end{tabular}
\end{table}

\subsection{Result Comparison (San Raphael Hospital)}
Compared to similar articles, comparisons have been made due to the general public of San Rafael Hospital. The result obtained in \cite{r43}, which is obtained with two procedures, standard with 88\% accuracy, 86\% sensitivity, and 91\% specificity, and the three-way version (a model that abstains from prediction when the confidence score is below 75\%) with 93\% accuracy, 92\% sensitivity, and 94\% specificity. In, several methods implemented on the data, the best outcome has been gained with sensitivity between 92\% and 95\%, and accuracy between 82\% and 86\%. It is worth noting in \cite{r37} the number of used features is more than in our study. Besides, other research related to our study with different datasets has been done. In \cite{r38}, by the routine blood tests of 6635 patients of BMI hospital, the results include 84\% recall, 84\% precision, and 84\% F1-score have been obtained. Our study with 95\% sensitivity, 94\% specificity, and 94,1\% accuracy, offers acceptable percentages compared to other methods.

\subsection{Radiographic Image Processing}
\subsubsection{Lung Segmentation}
Using the image segmentation dataset and the deep structure of U-Net, lung lobes can be separated from other organs in the image. By following this idea, the classifier structure will be able to make decisions based on the lung organ only, and other body organs will not be affected.

\begin{figure}
  \centering
  \includegraphics[width=13.5cm, height=6.5cm]{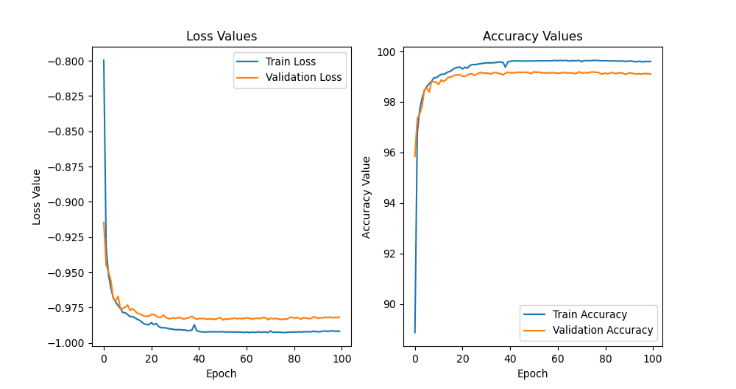} 
  \caption{Training and validation learning curves.}
  \label{fig8}
\end{figure}

In the deep structure of U-Net, the RELU functions are considered, but the convolutional layer at the end of the model, which has a one-to-one filter, uses the sigmoid activation function. The cost function is based on the Jaccard criterion and Adam's optimization algorithm is used to update the structure. 5\% of the data set is considered for validation, 10\% for testing and the remaining 85\% for the training process in groups with eight samples and 100 epochs. By applying the stated approach, the results will be checked in the community. The results of the training and validation process are displayed according to Figure \ref{fig8}.

\begin{figure}
  \centering
  \includegraphics[width=14.5cm, height=6.5cm]{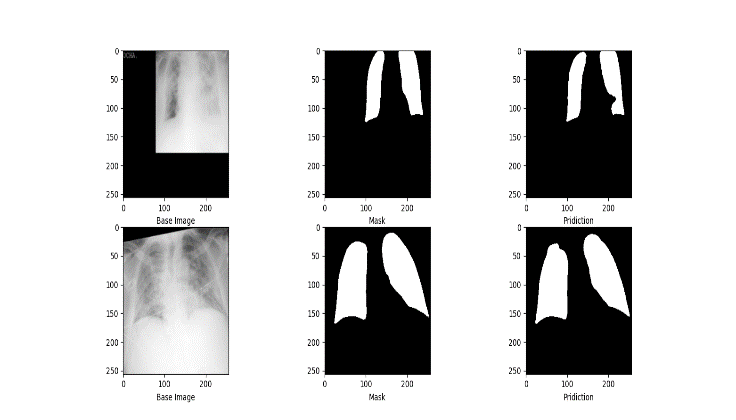} 
  \caption{Data samples for simulation schematic.}
  \label{fig9}
\end{figure}

The schematic of the simulation performed for two test data samples can be seen as described in Figure \ref{fig9}.

Each example displays the original image, the corresponding mask, and the segmented image as the output of the structure in order from the left side. The segmented image will be applied as input to the classifier structure. Next, the structure of the classifier and its results will be examined.

\subsubsection{Image Classification}
\begin{figure}
  \centering
  \includegraphics[width=13.5cm, height=6.5cm]{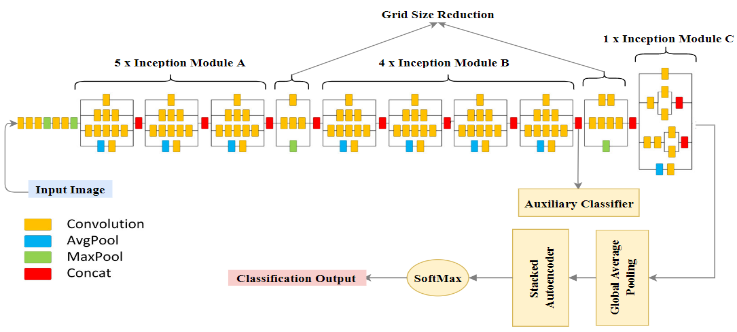} 
  \caption{Training and validation learning curves.}
  \label{fig10}
\end{figure}

The basic structure criterion used is taken from the initial deep convolution network-version three. In the initial network architecture version three, there are two initial modules expressed at the end of the structure. In the proposed model, one of the mentioned modules is removed and a stacked autoencoder is installed at the end of the convolutional model. Figure \ref{fig10} shows the schematic of the proposed model.

\subsubsection{Structure and Specification of Autoencoder}
The input to the vector stack autoencoder is 2048-in-1. The details of the coding layers in the stack autoencoder used in the proposed structure are described in Table \ref{table8} and \ref{table9}.

\begin{table}[ht]
\centering
\caption{Structural specification of stack autoencoder}
\label{table8}
\begin{tabular}{c|ccc}
                    & First layer             & Second layer            & Third layer             \\ \hline
Number of Neurons   & 512                     & 128                     & 32                      \\
Activation Function & Exponential Linear Unit & Exponential Linear Unit & Exponential linear unit
\end{tabular}
\end{table}

\begin{table}[ht]
\centering
\caption{Specification of stack autoencoder training process}
\label{table9}
\begin{tabular}{cccc}
Loss Function & Optimizer & Epoch & Batch-Size \\ \hline
MSE           & Adam      & 150   & 16 
\end{tabular}
\end{table}

At first, the initial deep convolutional structure is trained by removing a module in Figure \ref{fig10} along with the SoftMax activator function, and then its output is saved as input for training the stack autoencoder. The stack autoencoder is initially trained independently by the mentioned data set without a supervisor. 80\% of the data set is considered for raining, 15\% for testing, and 5\% for validation. 
The value of the cost function resulting from the network test data is 0.0017.

\subsubsection{Classifier Design Process}
After completing the training process of the stacked autoencoder network, the initial reduced convolutional deep neural network architecture shown in Figure \ref{fig10} is placed in series with the stacked autoencoder, and the structure set is under global training with 80\% The dataset is used for training and 20\% for testing and validation. Note that the input data set to the classifier includes segmented images without pre-processing by U-Net architecture. The details of the learning process are specified in Table \ref{table10}.

\begin{table}[ht]
\centering
\caption{Specification of stack autoencoder training process}
\label{table10}
\begin{tabular}{cccc}
Loss Function & Optimizer & Epoch & Batch-Size \\ \hline
Sparse Categorical Cross Entropy           & Adam      & 80   & 8 
\end{tabular}
\end{table}

\begin{figure}
  \centering
  \includegraphics[width=15.5cm, height=6.5cm]{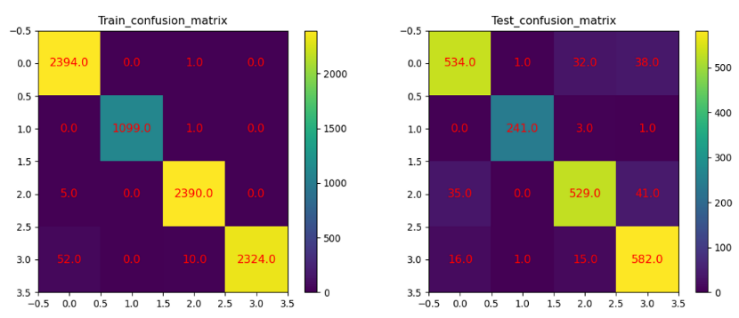} 
  \caption{Confusion matrices of the train and test sets.}
  \label{fig11}
\end{figure}

It should be noted that the activation functions in the convolution layers are RELU. After the completion of the training process, the obtained results will be evaluated. The confusion matrices of the training and test data sets are presented according to Figure \ref{fig11}.
Element 1-1 in the confusion matrices of Figure \ref{fig11} is related to the number of correctly recognized individuals of the normal class. Element 2-2 is related to the number of correctly diagnosed patients with viral pneumonia. Elements 3-3 and 4-4 also correspond to the number of correct diagnoses of patients related to Grand opacities and COVID-19, respectively.

For applying the test data set with the trained structure, the values of accuracy and cost function have reached 91.1\% and 0.472, respectively. To evaluate more accurately, the results obtained by the proposed structure are compared with the numbers obtained from the modified architectures. In addition to the proposed structure, in another architecture, instead of the stacked autoencoder, a multilayer neural network with the structural characteristics of the table is placed. Then, in another approach, the output of the convolutional structure is directly applied to the SoftMax activation function. The comparison of the stated structures is according to Table \ref{table11}.

\begin{table}[ht]
\centering
\caption{Comparison results of different structures}
\label{table11}
\begin{tabular}{c|cc}
Structure                                                        & Accuracy & Loss  \\ \hline
Reduced Inception V3+SAE+SoftMax (Proposed Model)            & 91.1\%   & 0.472 \\
Reduced Inception V3+SoftMax                                 & 88.9\%   & 0.555 \\
Reduced Inception V3+Fully Connected Neural Network +SoftMax & 89.0\%   & 0.511
\end{tabular}
\end{table}

In addition to comparing the results by making changes in the structure, the input data set in three modes without pre-processing, histogram equalization, and adaptive histogram equalization pre-processing with limited contrast, is applied to the proposed structure in Figure \ref{fig10} and their comparison is as described in Table \ref{table12}.

\begin{table}[ht]
\centering
\caption{Comparison results of different pre-processing}
\label{table12}
\begin{tabular}{c|cc}
Pre-processing                                                        & Accuracy & Loss  \\ \hline
Original Image            & 91.1\%   & 0.472 \\
Histogram Equalization Mapping                                & 88.2\%   & 0.672 \\
Contrast Limited Adaptive Histogram Equalization Mapping  & 88.7\%   & 0.557
\end{tabular}
\end{table}

\subsubsection{Visualization Result}
\begin{figure}
  \centering
  \includegraphics[width=13.5cm, height=6.5cm]{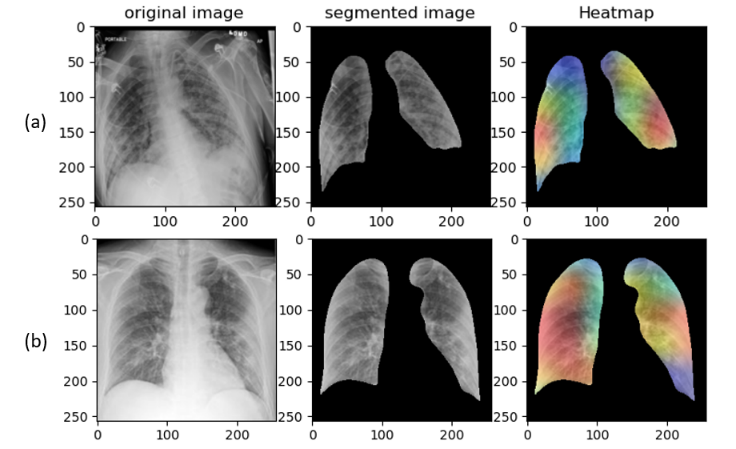} 
  \caption{Confusion matrices of the train and test sets.}
  \label{fig12}
\end{figure}

In this section, visualization results using the Grad-CAM method will be presented. The application of the visualization process can be seen in helping specialists to observe the abnormal areas of the lung, including the view of ground glass opacities and the phenomenon of consolidation. Figure \ref{fig12} shows two examples of lung images along with their heat map.
According to Figure \ref{fig12}, the pixels are displayed in different colors based on the degree of importance, and extreme amounts of infection and abnormal tissues are shown in red, orange, and yellow colors.

\section{Discussion}
COVID-19 viruses have rapidly spread worldwide, posing a significant global health threat with a high mortality rate \cite{r43}. The quick and accurate diagnosis of this disease is crucial to curb its further spread and initiate timely treatment for patients. Considering the constraints of time, cost, and the virus's mutation, it is essential to develop a reliable and resilient diagnostic method. In our study, we employed both blood test parameters and image processing techniques to enhance the accuracy and efficiency of COVID-19 diagnosis.

For blood test analysis, we utilized two datasets: the blood test dataset from San Raphael Hospital. These datasets contained various blood parameters, including CRP, Lymphocyte, Platelet, W.B.C, LDH, AST, ALT, RBC, Urea, and Neutrophils. By applying different classification methods such as KNN, Random Forest, MLP, RBF Neural Network, Fuzzy Network, Ensemble Method, and SVM, we evaluated the performance of each method in disease detection. The results showed that the Ensemble Method achieved the highest performance with an F1-score of 94.69\% on the blood dataset from Raphael Hospital.

In addition to blood test analysis, our study incorporated image processing techniques, particularly the U-Net deep structure for lung segmentation. By effectively separating lung lobes from other organs in radiography images, we ensured that the subsequent classification model focused exclusively on the lung region, minimizing the influence of irrelevant factors and improving diagnostic accuracy.

For image classification, we developed a modified architecture by integrating a stacked autoencoder with the deep convolutional network. This approach enhanced the model's ability to extract relevant features from the segmented lung images, leading to improved classification performance. The ensemble method demonstrated the best overall performance, achieving an accuracy of 91.1\% on the test data.

The combination of blood test parameter analysis and image processing techniques provides a comprehensive diagnostic approach for COVID-19. By leveraging both these methods, we obtained significant advancements in disease detection accuracy, reducing the likelihood of false positives or negatives. This integrated approach offers several advantages, including cost-effectiveness, shorter diagnosis time, and the potential for implementation in areas with limited healthcare resources.

Further research can be conducted to enhance the diagnostic capabilities of the proposed methods. Increasing the size of the datasets would improve the robustness and generalizability of the models, allowing for more accurate disease diagnosis. Additionally, consulting medical experts or employing feature selection techniques can expand the number of relevant parameters, making the trained models more comprehensive. Incorporating deep learning algorithms and expanding the dataset would enable the identification of hidden features in the data, leading to even higher accuracy in disease detection.
\section{Conclusion}
In conclusion, our research highlights the significant role that artificial intelligence, machine learning, and image processing can play in enhancing the diagnosis of COVID-19, making it more accurate and efficient. By integrating blood test analysis and advanced image processing, we have formulated a dependable and economical diagnostic solution beneficial for healthcare settings, especially those with limited resources. Our approach, which includes the analysis of various blood parameters and the application of sophisticated classification and image processing techniques, has led to notable improvements in identifying the disease. The Ensemble Method, in particular, stood out for its effectiveness across different datasets.

Moreover, by employing cutting-edge image processing to analyze lung images specifically, we've managed to increase the accuracy of our diagnoses further. This progress underscores the vast potential of leveraging AI and machine learning in medical diagnostics. Looking ahead, expanding our dataset, integrating expert insights, and exploring deeper learning algorithms promise to enhance these diagnostic methods' accuracy and applicability even more. This forward momentum paves the way for more sophisticated medical AI tools and promises to make high-quality healthcare diagnostics more accessible and affordable globally.

\section*{Declarations}
Conflict of interest The authors declare that there is no conflict of interest regarding the publication of this paper
\section*{Acknowledgment}
We thank the experts at San Rafael Hospitals for providing the data, and we appreciate the suggestions of the anonymous referees.

{\small
\bibliographystyle{unsrt} 
\bibliography{egbib}
}

\end{document}